# Molecular filtration by reduced graphene oxide sandwiched between wood sheets

Eguono W. Omagamre[a], Mahdi J. Fotouhi[a], Tyler Washington[a], Chinedu Ahuchaogu[a], Yaganeh Mansurian[a], Sudarshan Kundu[b] and Kausik S Das [a, *]

[a]Department of Natural Sciences, University of Maryland Eastern Shore, Princess Anne, MD, USA,  [b]Kent State University, Kent, Ohio, USA
[*]kdas@umes.edu

This study investigates the prospect of using reduced graphene oxide encapsulated between thin wood membranes as a composite filter for water filtration. Graphene oxide (GO) with a volume range of 0.2 to 0.5 mL was deposited on a 95 mm$^2$ area of thin sheets of Balsa wood with a thickness range of 2.20 - 3.50 mm. The dried GO deposits were laser treated in reduced oxygen environment to prepare a reduced graphene oxide-wood composite (rGO-wood). The rGO substrate of the composite was sandwiched between 2 wood sheets which were glued using polydimethylsiloxane (PDMS) to give a stable membrane. Fabricated membranes improved the Allura red rejection from 8.29 ± 0.69 % in the controls (sandwiched wood sheets without rGO) to 93.14 ± 1.31 % in the 1 mL total rGO composites. The ratio of percentage permeace drop of the treatments over the controls ranged from 62.8 ± 4.3 to 85.4 ± 5.2 % from the 0.4 to 1 mL rGO composites respectively. The correlation coefficient (r) between the GO volume and rejection was 0.94 suggesting that the GO treatments provided some control over the transport of molecules through the membrane. The correlation coefficient between wood thickness, and Allura red rejection shows that the wood sheets did not influence the rejection. The permeance was weakly affected by the membrane thickness (r = -0.46). An 93% improvement in rejection performance of the composite membranes over the controls was observed. The overall data suggests that the wood-rGO composite has potential for water filtration applications.

## Introduction

The need to reduce human exposure to many chemicals of emerging concern in water has been a driving force for research into inexpensive, effective, energy efficient and environmentally friendly water treatment systems. Many of these studies involve evaluating the filtration efficiency and stability of filters with nanopores and nanochannels(1, 2) or materials with large surface area and chemically active sites(3, 4). For instance, Burgmayer and Murray(5) studied the transport and rejection properties of an electrically controlled gold grid channel coated with a conducting polymer over three decades ago. Water transport and filtration through carbon nanotubes, nano-sized pores fabricated in monolayer graphene, as well as with multilayer graphene oxide (GO) sheets have been studied(6–8). Ions with hydration shells greater than 13Å was rejected by 2 D slits devices fabricated from graphite(9). The use of 2D materials as separation membrane in several researches have been focused on enforcing selectivity via stearic confinements(2, 5, 6). Properties required in membranes usable for filtration/adsorption operations include good water flux and permeability, high mechanical and chemical stability, resistance to fouling and consistent rejection(10). Graphene oxide membranes have been demonstrated to show these properties(11). Graphene oxide easily disperses in water and this makes it easy to fabricate into single and multilayer 2D sheets(12). The presence of carboxyl, hydroxyl and epoxy functional groups enable GO to form good aqueous suspension and ready to be fabricated into multilayer sheets(13). The low-frictional flow of a monolayer of water through 2D capillaries of GO sheets(6) which highlights the plausibility of high water flux through GO multilayer membranes in filtration applications has been demonstrated in a number of molecular simulations(14–16). These demonstrations correlate well with the postulation that obtaining completely dry GO is not practically possible given the hydrogen-bonding functional groups present on its surface(12, 17, 18).

The regular interlayer spacing and the charges on GO sheets are two properties that make GO membranes highly appealing towards water purification applications(19–21). The charged groups on GO have been shown to have affinity for a range of organic and inorganic water contaminants(22–28) and the angstrom order interlayer spacing have been reported to enforce selective transport(6, 21, 23–29). However, technical difficulties associated with the practical, largescale use of graphene-based filtration membranes arise due to their instability during application(30). Their use in filtration applications are limited by their effective interlayer spacing of ~9 Å due to their swelling and dispersity in water(29). This swelling has been attributed to ready bonding of water to the ionizable edge carboxylic functional groups on the sheets(31–33). Hence, cut-off for smaller molecules are difficult using as-is GO membranes. Attempts that have been exploited to reduce

swelling include partial reduction, covalent crosslinking and physical confinements(29, 34, 35)

To stabilize and confine GO sheets, Shin et al(13) immobilized 5 μm graphene oxide multilayer on polyethersulfone (PES) porous membrane with a thickness of ~130 μm and utilized it for water-ethanol separation. Abraham et al(29) utilized an epoxy stycast to encapsulate and stack multilayer graphene oxide sheets giving rise to a membrane with cut-off of ~9 – 6.4 Å. Zhang and other researchers(2) grafted ethylene glycol on the surface of GO to alter the distance between its layers and improve its rejection. Graphene oxide/polyacrylamide composite membranes with proper interlayer spacing and improved performance over pristine GO was prepared by Cheng et al.(36) via vacuum filtration. Polydopamine has been used as an interlaminar short-chain molecular bridge to covalently interlock GO laminates resulting in relatively stable membranes compared to pristine GO(37).

While these physical confinement strategies improved the stability of GO for filtration applications, the issue of environmental and health concerns of the membranes arise since the polymeric materials used for these confinements may leach into the water being purified and some of them are not readily biodegradable after the membranes serve their useful lives. In this study we explored the potential of immobilizing a multilayer of reduced GO between thin wood sheets to form a biodegradable composite. It is hypothesized that the reduction of the GO and its confinement between the thin wood sheets will provide reproducible stability for the GO membrane. We have confirmed that the porous structure of wood sheets supports good water flux for the fabricated composite membrane(38).

## MATERIALS AND METHODS

### 2.1 Fabrication of wood chips

Balsa wood block purchased from Midwest Products (Hobart, Indiana) was used for this experiment. Thin sheets were chipped off from the transverse sections of the wood block. The sheets were planed and thinned further by rubbing with a sandpaper grit to achieve a thickness range of 2.20 to 3.57 mm. Roughly square sheets of 1.5 x 1.5 cm were cut out of the smoothened sheets. The obtained sheets were repeatedly rinsed in deionized water until a largely clear water filtrate was observed. Thereafter they were placed in deionized water and sonicated for 30 minutes to aid the removal of dust particles clogged in the pores. The sheets were thereafter dried at 40°C overnight.

### 2.2 Graphene Oxide Treatment

Prior to deposition of GO on the surface of the chips, an 11 mm circle was marked on the wood chip surface using laser. Thereafter, all areas outside the circle was covered with polydimethylsiloxane (PDMS) and allowed to dry, creating a hydrophobic surface outside the circle. Graphene oxide was then pipetted into the circular interior and allowed to dry at 40 °C overnight in an oven. Graphene volumes of 200 to 500 μL on individual wood sheets were investigated in this study.

### 2.3 Graphene oxide reduction, Characterization and membrane fabrication

The wood sheets containing dried graphene oxide were laser treated under reduced oxygen conditions. Briefly, the dried wood-GO membrane was attached to the inner surface of a glass petri dish with the GO surface in contact with the dish. The petri dish was then placed on a vacuum stand, and the pressure in the dish was evacuated to 2.2 Torr. Under this condition, a laser beam with a wavelength of 405 nm set to a 35% power of 317 milliwatts power rating was used to scribe the GO surface for about 30 seconds.

A Fourier transform infrared spectroscopy (FTIR; Thermo Fisher Jessup, MD, USA) was used to investigate the reduction of oxygenated functional groups. A raman spectroscope (Thermo Fisher, Jessup, MD, USA) with a 532 nm laser and power of 1 mW was used to evaluate the quality of rGO produced from GO. The surface morphology of the membranes as well as the thickness of the GO layers were studied using a scanning electron microscope (SEM; NanoScience, Phoenix, AZ, USA)

Two pieces of the wood-rGO membrane (with the same GO volume treatment) were then sandwiched and glued together by applying PDMS over the non-GO areas. Pressure was applied over the sandwiched sheets overnight at room temperature to give a stable membrane with fully immobilized rGO between the wood sheets.

### 2.4 Flowrate and diffusion experiment

Fabricated composite membranes were placed in deionized water and allowed to sit for an hour to wet their pores after which the thicknesses of the wet membranes were measured. The rejection of Allura red molecules in solution through the membranes was then carried out using Side-Bi-Side PermeGear diffusion cells (Hellertown, PA, USA). The cells have a 6 mm orifice which allows water flux through the membrane. A pressure difference of 0.5 bar was used to drive the feed solution of Allura red dye (10Å molecular diameter) through the membrane. The absorbance of the feed and the permeate solutions were determined using a UV-Visible spectrophotometer (Brea, CA, USA). The volume of permeate solution was also measured over time to determine the hydraulic flux of the membranes. Figure 1 summarizes the steps involved in the membrane fabrication.

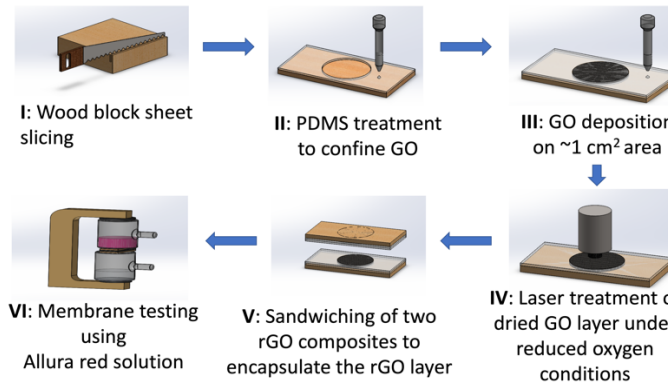

**Fig. 1**: *Schematics illustrating the fabrication and diffusion experimentation of the wood-rGO composite membrane*

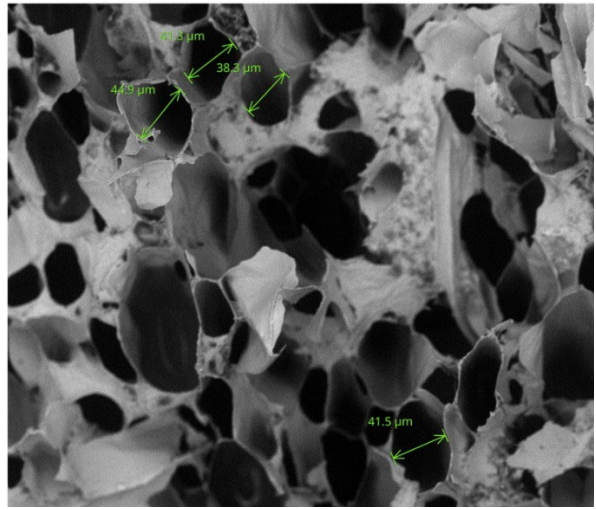

**Fig. 2**: *SEM image of the surface of thin wood sheets prior to GO treatment*

## RESULT AND DISCUSSIONS

*3.1     Pore Characterization of Wood Membrane*

The obtained SEM images from the wood sheets show most of the pore diameters within the 40 μm range as seen in Figure 2. Bigger vessels, which are the axial parenchyma cells, with diameters in the 200 μm range were sparsely distributed in the sheets. Wiedenhoeft(38) reported that vessels in hardwoods may have diameter that range from 30 – 300 μm. Overall, the cutting and filing of the wood sheets produced some collapsed and blocked pores. The pore sizes highlight why the wood membranes may not be effective against angstrom order molecules.

*3.2     Reduced graphene oxide membrane characterization*

The functional groups on the GO were evaluated before and after laser scribing. The graphene oxide (GO) film on the wood membrane prior to scribing (Fig. 3a) shows a broad O-H stretch from 2500 – 3600 cm$^{-1}$. Hydroxyl stretch vibration have been shown to be present around this region in GO(39, 40). The peaks at 1045 and 1710 cm$^{-1}$ corresponds to C–O (epoxy) and C=O (carbonyl) respectively as also observed in the GO spectra obtained by Rao and others(39–42).

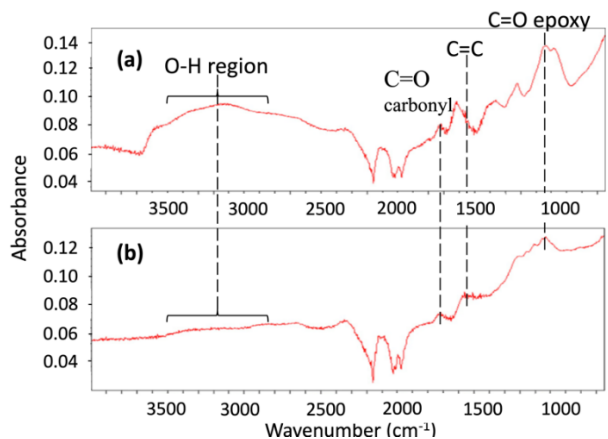

**Fig. 3**: *FTIR of **(a)** GO deposit on wood membrane **(b)** reduced GO on wood membrane*

After laser treatment, observable changes occurred with respect to the oxygenated functional groups in the GO. The reduced GO (rGO) showed a disappearance in the O-H stretch present in the GO as seen in Fig. 3b. The removal of the O-H stretch is one of the modifications looked for in the evaluation of the reduction of GO to graphene(41). A distinct C=C stretch at 1550 cm$^{-1}$ is attributed to an aromatic unoxidized sp$^2$ hybridized bond which is also a characteristic bond in pristine graphene(43–45). The presence, albeit with reduced intensities, of the epoxy and carbonyl functional groups at 1045 and 1710 cm$^{-1}$ respectively however indicates that the reduction of GO to graphene was incomplete(40, 42).

The raman spectra shows the characteristic D, G and 2D bands for graphene oxide (Fig. 4). The D vibrations are shown at 1348.94 cm$^{-1}$ and 1341.22 cm$^{-1}$ for the GO and rGO respectively. The D-band is reflective of defective/disordered sheets or edges of sheets arising from coexistence of sp$^2$ and sp$^3$ hybridized carbon domains(46, 47). This band is usually absent in the raman spectra for graphite because of its more ordered structure i.e. highly crystalline, compared to exfoliated GO and corresponding rGO. Similar frequencies for the D-band has been reported by other researchers(46, 48, 49). The $I_D/I_G$ ratio increased from 0.99 to 1.28 on reducing the GO to rGO which is an indication of increased defect/disorder on the sheets arising from the thermal reduction process(50).

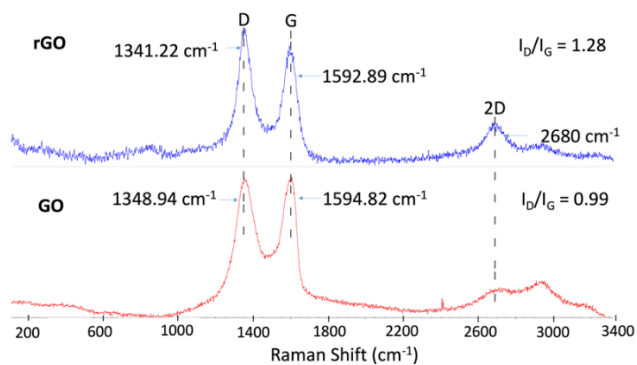

**Fig. 4:** *Raman spectra of rGO deposit and GO on wood membrane*

The 2D peak has been shown to be very sensitive to the pi band in the electronic structure of graphene and is used to evaluate the stacking of graphene multilayers(46, 48, 50). The 2D band has a higher and well defined peak in the rGO membrane suggesting that the reduction restored some of the C=C sp$^2$ configuration that was lost due to oxidation of graphene to GO (during the GO fabrication process which is out of the scope of this paper) giving sp$^3$ hybridized carbon. The sp$^2$ configuration thus allows for better stacking of the graphene layers in rGO compared to GO(48). Further, there appears to be a shift of the 2D peak to higher wavenumbers in the GO (2950 cm$^{-1}$) compared to rGO (2680 cm$^{-1}$) which is a reflection of the presence of more oxidized functional groups in the GO(50).

It was observed that the GO deposit prior to laser treatment dissolved readily in water even after significant drying on the wood sheet. This phenomenon is believed to arise from the electrostatic repulsion of the negative charges generated on the stacks after hydration(30). However, after scribing, the rGO remained much more attached to the wood membranes even when immersed in water for over 48 hours. This stability may be connected with the reduced formation of negative charges due to a reduction of most of the hydroxy and other hydrophilic groups during the scribing. The rGO showed significant hydrophobic characteristics when immersed in water. This hydrophobicity suggests that the rGO formed less hydrogen bonding with water than GO. The basal plane of the GO, which contains essentially networks of polyaromatic benzene rings(31, 32, 51, 52) that should now be predominantly at play, may be responsible for this observation.

The SEM images of the GO membranes before and after treatment are shown in Fig. 5. The rGO (5b) shows a wrinkled, flaked, and more crystalline morphology compared to the GO (5a). A similar transformation in surface morphology was observed in the reduction of graphene oxide using green tea extracts by Sykam and others(42). The flaking and crumpling in rGO has been attributed to the thermodynamic stabilization of its 2-dimensional morphology being formed via the reduction. Meyer et. al.(53) indicated that this sort of curling is intrinsic to 2-dimensional membranes such as graphene. These wrinkles are responsible for the disorder and the loss of 3-dimensional ordering in the stacked sheets as explained earlier leading to a deviation from the sp$^2$/planar character expected for graphene(46).

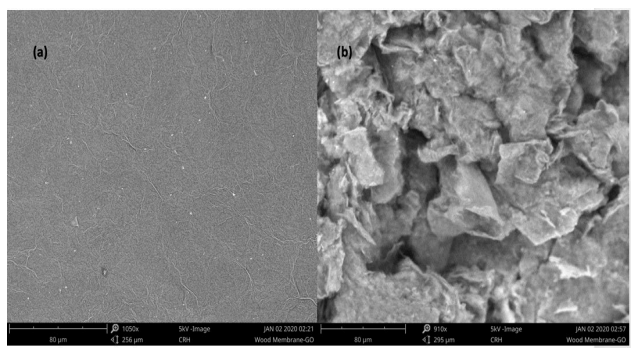

**Fig 5:** *SEM image of (**a**) GO deposit on wood membrane (**b**) reduced GO on wood membrane*

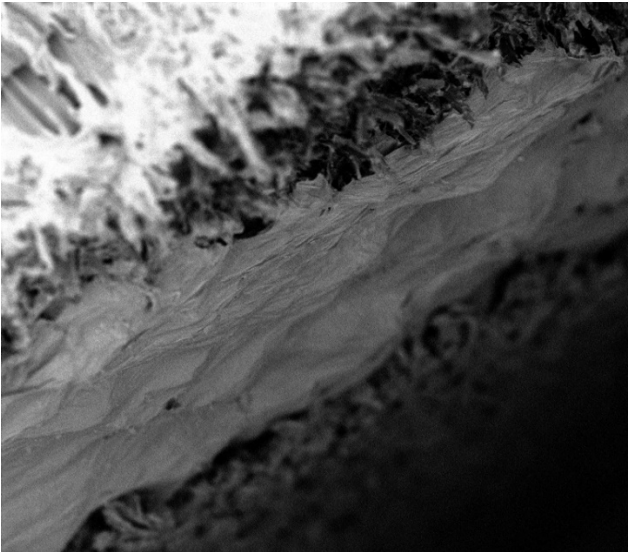

**Fig 6:** *SEM image of cross-sectional cut of membrane to show the rGO stacks between the wood sheets.*

*3.3       Water Flux of Membrane*

A high permeance is desired in filtration membranes as it reflects reduced pressure drop, improved water flux, thus requiring lesser energy for filtration. The permeance of the composite membranes were computed from the permeate flowrates through the membranes using equation (1):

$$\frac{\text{volume flow rate across membrane (L h}^{-1})}{\text{Membrane area (m}^2) \cdot \text{Pressure difference (bar)}} \quad \ldots\ldots\ldots (1)$$

The hydrodynamic conductivity (k) [see equation 2] was also calculated for the control membranes to compare with values reported by Boutilier and others(56) for their pine wood filter system. It can be expressed as;

$$k = \frac{Q\,L}{A\,\Delta P} \ldots\ldots\ldots\ldots\ldots (2)$$

where Q represents the rate of water flow in m$^3$/s through the membrane of thickness L (m) and cross-sectional area of A (m$^2$) across a pressure drop of $\Delta$P pascal. The Hydraulic conductivity for the control membranes ranged from $1.19 \times 10^{-7}$ - $4.61 \times 10^{-7}$ m$^2$ s$^{-1}$ Pa$^{-1}$. This range was higher than the approximate $5 - 6 \times 10^{-10}$ m$^2$ s$^{-1}$ Pa$^{-1}$ reported by Boutilier and others (56) where they used a 1-in section of pine branch with a 1 cm diameter for filtration. The reduced thickness of the wood membranes and possibly the wood type in this study seemed to play a role in the improved hydraulic conductivity compared to that study.

For further comparison of the water flux of composite membrane with other graphene oxide-based membranes, the permeance is discussed (Fig. 7 and 8). With the sandwiching of the reduced graphene oxide between the wood membranes in this study, the obtained permeance ranged from 195.0 $\pm$ 22.2 L m$^{-2}$ bar$^{-1}$ h$^{-1}$ in the 400 µL to 76.5 $\pm$ 27.4 L m$^{-2}$ bar$^{-1}$ h$^{-1}$ in the 1000 µL total GO treatments (Fig 7).

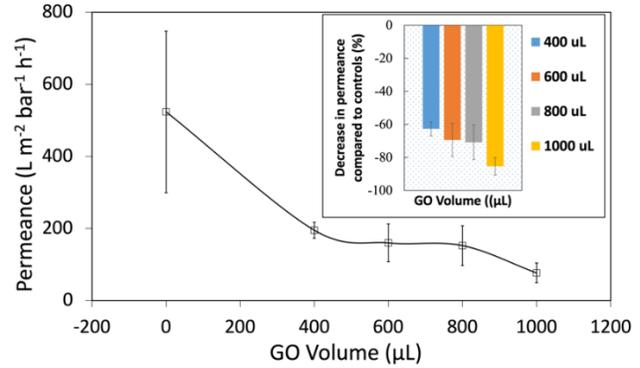

**Fig 7:** *Effect of graphene oxide treatment volume on the permeance of the membranes. Inset shows percentage decrease in permeance of the treatments over the controls. The error bars represent the standard error of the means (n =3).*

These conductivities reflect percentage permeance drop ranging from 62.8% to 85.4% respectively compared to the membranes without rGO. The percentage drop was fairly constant for 600 and 800 µL treatments with values of 69.5 and 70.9% respectively. The obtained permeance in this study were higher than the approximately 0.5 – 1.0 L m$^{-2}$ bar$^{-1}$h$^{-1}$ reported by Abraham et. al(29) where they encapsulated layers of graphene oxide using Stycast epoxy polymer. The reduced permeance seem to be a reflection of the effective distance between the layers as molecules of 7.9 Å were confined in that study. 100% filtration was not achieved with the ≈10 Å Allura red molecules.

The impact of the reduction and immobilization of the GO on the reduction in the amount of hydration and swelling is seen in the permeance range of 399.04 L m$^{-2}$ bar$^{-1}$ h$^{-1}$ obtained by Cheng and others(36) where they used as-is GO as compared to an optimal 76.5 L m$^{-2}$ bar$^{-1}$ h$^{-1}$ obtained in this study for rGO. This optimal flux value is close to the flux/permeance of 85.85 L m$^{-2}$ bar$^{-1}$ h$^{-1}$ for rGO membrane fabricated by the same researchers. Our flux data may highlight some of the positives of combining reduction and immobilization technique on the same membrane. Han and others(54) obtained a maximum flux of 51 L m$^{-2}$ bar$^{-1}$ h$^{-1}$ from their fabricated base-refluxed reduced GO (brGO) membrane. Their membranes may have undergone more reduction than ours since oxygen containing functional groups is believed to form a layered capillary structure that expands the GO layers that eventually allows fast transport of water molecules over hydrophobic regions of the surface(55).

Thus, a reduction in the numbers of these hydrophilic groups may translate to less swelling and eventually slower flux of water molecules.

Evaluation of the impact of the thickness of the wood sheet on the rate of water transport was determined by assessing the correlation between the wood membrane thickness and the water flux of the membranes. As observed from Figure 8, a slight correlation (r =-0.46) was observed suggesting that the increase in thickness of the fabricated membranes (which is a combination of the thickness of the wood sheets and GO volume) has an inverse relationship with the permeance.

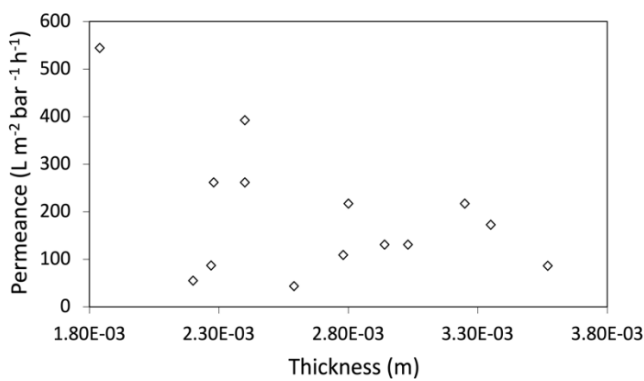

**Fig 8**: *Effect of Membrane thickness on permeance of the fabricated membranes*

This suggests that by reducing the thickness of the wood sheets used for encapsulating the rGO, minimal improvements in the permeance may be obtained. The results also suggests that the rGO layers influenced the pearmeance drop in the composites more than the wood sheets as a correlation coefficient of -0.67 was obtained between the GO volume and permeance. Abraham et al(29) observed that a decrease in membrane thickness from 5 μm to 1 μm improved the water flux from 0.6 to 2.5 Lm$^{-2}$h$^{-1}$. This indicates that increasing GO layers impacts permeance.

### 3.4 Allura Red Rejection Studies

The rejection of the approximately 7 x 10$^{-6}$ M solution of Allura red was used to evaluate the selectivity of the membranes via vacuum filtration. The obtained results are captured as Fig. 9.

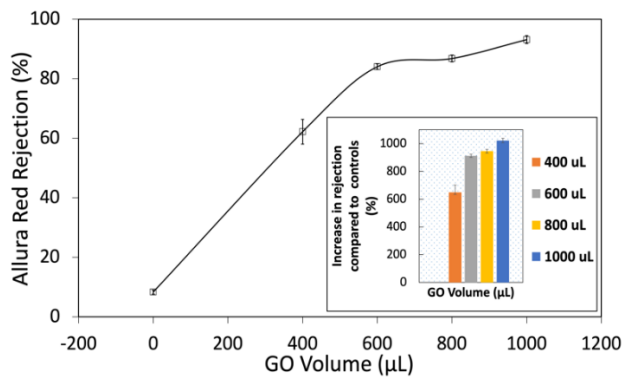

**Fig 9:** *Effect of Graphene Oxide treatment volume on percentage rejection of Allura red molecules. Inset shows percentage improvement in Allura red rejection in the treatments over controls. The error bars represent the standard error of the means (n =3).*

The mean percentage rejection of the wood membranes without rGO treatments was 8.29 ± 0.69 %. The composite rGo membranes showed mean rejection that ranged from 62.2 ± 4.1 to 93.1 ± 1.3 % respectively for the 400 to 1000 μL treatments. This suggests that the GO layers play significant role in the effective filtration of the fabricated membranes. In the study by Boutlier et al.(56) that explored the xylem and piths in wood membrane for filtration, molecules smaller than ≈ 80 nm could not be filtered out. This may also explain the poor filtration of the ≈ 1 nm Allura red molecules (mwt. 496.02 g/mol) by the wood membranes without rGO treatments that served as controls in this study and the subsequent improvements with the rGO treatments. The rejection improvements with increase in GO treatment volume may be due to increased rGO layers which reduces the chances of permeation of Allura red molecules through cracks and imperfections in some of the rGO layers. Hu and Mi (21) also observed that as the number of deposited GO layers increases, the membrane rejection performance also increased. The improvements with more volume may also be as a result of more surface adsorption of the allura red to the increased number of rGO sheets/layers via hydrophobic and electrostatic interactions. Ultra violet-visible characterization of the surface scrappings of the rGO membranes before and after filtration of Allura red molecule is shown in Fig. 10. The presence of aromatic peaks at 253 nm in both the Allura red and the rGO (after filtration) spectra suggests the rGO retained the Allura red molecules after filtration.

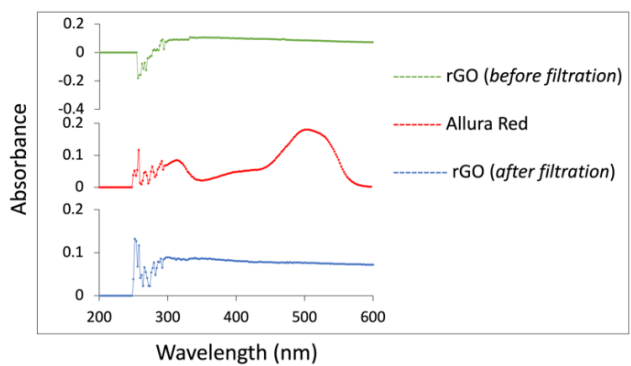

**Fig. 11**: *Characterization of the rGO surface before and after filtration of Allura red over a UV-VIS range of 200 - 600 nm. Obtained spectra of Allura red is also shown.*

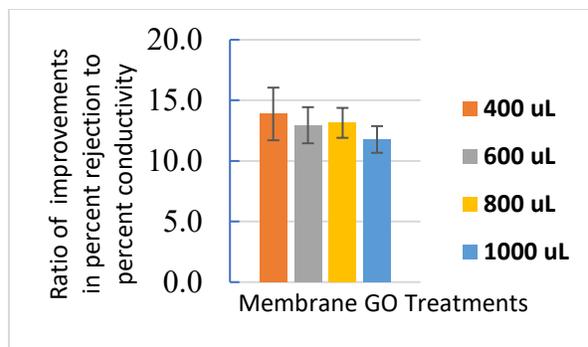

**Fig. 11**: *Performance evaluation of fabricated membranes in comparison to controls. The error bars represent the standard error of the means (n =3)*

The rejection error percent of replicates for each GO treatment is under 1.5% except for the 400 μL treatments that is 4.1%. These low error bars suggest the reproducibility of the membrane fabrication method used. The effect of the thickness of the wood sheets on the rejection performance of the composite membranes was assessed and no correlation was observed over the combined thickness variation of 0.48 mm of all membranes used in the study. This suggests that the thickness of the wood sheets did not play a significant role on the rejection performance of the composite membranes. It may also be inferred that reducing the wood sheet thickness to improve the permeance may not affect the rejection efficiency to an extent.

The overall performance of the fabricated membranes in relation to the controls was determined by evaluating the ratio of the percent improvements of the Allura red rejection to the drop in hydraulic conductivity (absolute value). This value suggests that the fabricated composite membranes showed an approximately 13-fold better performance over the membranes without rGO (Fig 10). Slight variation was observed in this perfomance across the composite membranes with a standard deviation of 0.87 across the membranes suggesting a consistency in performance accompanied by slight drop with increasing GO treatment, which is expected.

## Conclusions

In this study the potential of encapsulated reduced graphene oxide stacks between wood sheets to give stable, reproducible and composite membranes was explored. It was hypothesized that the xylem pores of the wood sheets would serve as transport channels for water molecules travelling through the graphene layers. Obtained data suggests that the reduced-graphene oxide layers exerted consistent filtration/adsorption of Allura red molecules. Rejection as high as 93% was observed in the composite membranes. While there was a decrease in permeance with GO treatments, obtained values were in the range of data from other studies. The data from the study suggests that encapsulated reduced-graphene oxide between wood sheets produces stable membranes with potential for water filtration via sieving and/or adsorption of contaminant molecules. Further studies are ongoing to understand the mechanism of filtration and the relationship between graphene volume treatment and molecular size cut-off.

## Conflicts of interest

There are no conflicts to declare

## Acknowledgements

This research is supported by MSEIP CCEM (Award # P120A170068).

## Notes and references

‡ Footnotes relating to the main text should appear here. These might include comments relevant to but not central

to the matter under discussion, limited experimental and spectral data, and crystallographic data.

1. Liu G-f, Huang L-j, Wang Y-x et al. Preparation of a graphene/silver hybrid membrane as a new nanofiltration membrane. RSC advances. 2017;7:49159-49165.

2. Zhang Y, Huang L-j, Wang Y-x et al. The preparation and study of ethylene glycol-modified graphene oxide membranes for water purification. Polymers. 2019;11:188.

3. Saleh NB, Khalid A, Tian Y et al. Removal of poly-and per-fluoroalkyl substances from aqueous systems by nano-enabled water treatment strategies. Environmental Science: Water Research & Technology. 2019;5:198-208.

4. Zhao C, Fan J, Chen D, Xu Y, Wang T. Microfluidics-generated graphene oxide microspheres and their application to removal of perfluorooctane sulfonate from polluted water. Nano Research. 2016;9:866-875.

5. Burgmayer P, Murray RW. An ion gate membrane: electrochemical control of ion permeability through a membrane with an embedded electrode. Journal of the American Chemical Society. 1982;104:6139-6140.

6. Joshi RK, Carbone P, Wang FC et al. Precise and ultrafast molecular sieving through graphene oxide membranes. science. 2014;343:752-754.

7. Kidambi PR, Jang D, Idrobo J et al. Nanoporous atomically thin graphene membranes for desalting and dialysis applications. Advanced Materials. 2017;29:1700277.

8. Secchi E, Marbach S, Niguès A, Stein D, Siria A, Bocquet L. Massive radius-dependent flow slippage in carbon nanotubes. Nature. 2016;537:210.

9. Esfandiar A, Radha B, Wang FC et al. Size effect in ion transport through angstrom-scale slits. Science. 2017;358:511-513.

10. Gogotsi Y. Moving ions confined between graphene sheets. Nature nanotechnology. 2018;13:625.

11. Mahmoud KA, Mansoor B, Mansour A, Khraisheh M. Functional graphene nanosheets: The next generation membranes for water desalination. Desalination. 2015;356:208-225.

12. Daio T, Bayer T, Ikuta T et al. In-situ ESEM and EELS observation of water uptake and ice formation in multilayer graphene oxide. Scientific reports. 2015;5:11807.

13. Shin Y, Taufique MFN, Devanathan R et al. Highly selective supported Graphene oxide Membranes for Water-ethanol separation. Scientific reports. 2019;9:2251.

14. Kumar Kannam S, Todd BD, Hansen JS, Daivis PJ. Slip length of water on graphene: Limitations of non-equilibrium molecular dynamics simulations. The Journal of chemical physics. 2012;136:024705.

15. Marti J, Sala J, Guardia E. Molecular dynamics simulations of water confined in graphene nanochannels: From ambient to supercritical environments. Journal of Molecular Liquids. 2010;153:72-78.

16. Xiong W, Liu JZ, Ma M, Xu Z, Sheridan J, Zheng Q. Strain engineering water transport in graphene nanochannels. Physical Review E. 2011;84:056329.

17. Bi H, Yin K, Xie X et al. Ultrahigh humidity sensitivity of graphene oxide. Scientific reports. 2013;3:2714.

18. Szabó T, Berkesi O, Dékány I. DRIFT study of deuterium-exchanged graphite oxide. Carbon. 2005;43:3186-3189.

19. Goh PS, Ismail AF. Graphene-based nanomaterial: The state-of-the-art material for cutting edge desalination technology. Desalination. 2015;356:115-128.

20. Nghiem LD, Schafer AI, Elimelech M. Removal of natural hormones by nanofiltration: Measurement, Modelling, and Mechanisms. Environ Sci Technol. 2004;38, 6:1888-1896.

21. Hu M, Mi B. Enabling graphene oxide nanosheets as water separation membranes. Environmental science & technology. 2013;47:3715-3723.

22. Lakshmi J, Vasudevan S. Graphene—a promising material for removal of perchlorate (ClO 4−) from water. Environmental Science and Pollution Research. 2013;20:5114-5124.

23. Liu T, Li Y, Du Q et al. Adsorption of methylene blue from aqueous solution by graphene. Colloids and Surfaces B: Biointerfaces. 2012;90:197-203.

24. Madadrang CJ, Kim HY, Gao G et al. Adsorption behavior of EDTA-graphene oxide for Pb (II) removal. ACS applied materials & interfaces. 2012;4:1186-1193.

25. Romanchuk AY, Slesarev AS, Kalmykov SN, Kosynkin DV, Tour JM. Graphene oxide for effective radionuclide removal. Physical Chemistry Chemical Physics. 2013;15:2321-2327.

26. Wang X, Huang S, Zhu L, Tian X, Li S, Tang H. Correlation between the adsorption ability and reduction degree of graphene oxide and tuning of adsorption of phenolic compounds. Carbon. 2014;69:101-112.

27. Williams CD, Carbone P. Selective removal of technetium from water using graphene oxide membranes. Environmental science & technology. 2016;50:3875-3881.

28. Xu J, Wang L, Zhu Y. Decontamination of bisphenol A from aqueous solution by graphene adsorption. Langmuir. 2012;28:8418-8425.

29. Abraham J, Vasu KS, Williams CD et al. Tunable sieving of ions using graphene oxide membranes. Nature nanotechnology. 2017;12:546.

30. Yeh C-N, Raidongia K, Shao J, Yang Q-H, Huang J. On the origin of the stability of graphene oxide membranes in water. Nature chemistry. 2015;7:166.

31. Kim J, Cote LJ, Kim F, Yuan W, Shull KR, Huang J. Graphene oxide sheets at interfaces. Journal of the American Chemical Society. 2010;132:8180-8186.

32. Lerf A, He HY, Forster M, Klinowski J. A new structural model for graphite oxide. J Phys Chem B. 1998;102:4477-4482.

33. Li DM. S., Kaner, RB & Wallace, GG. Nature Nanotechnol. 2008;3:101-105.

34. Hung W-S, Tsou C-H, De Guzman M et al. Cross-linking with diamine monomers to prepare composite graphene oxide-framework membranes with varying d-spacing. Chemistry of Materials. 2014;26:2983-2990.

35. Liu H, Wang H, Zhang X. Facile fabrication of freestanding ultrathin reduced graphene oxide membranes for water purification. Advanced materials. 2015;27:249-254.

36. Cheng M-m, Huang L-j, Wang Y-x et al. Synthesis of graphene oxide/polyacrylamide composite membranes for organic dyes/water separation in water purification. Journal of Materials Science. 2019;54:252-264.

37. Zhang M, Mao Y, Liu G, Liu G, Fan Y, Jin W. Molecular bridges stabilize graphene oxide membranes in water. Angewandte Chemie International Edition. 2019

38. Wiedenhoeft A. Structure and function of wood. Wood handbook: wood as an engineering material: chapter 3 Centennial ed General technical report FPL; GTR-190 Madison, WI: US Dept of Agriculture, Forest Service, Forest Products Laboratory, 2010: p 31-318. 2010;190:3.1-3.18.

39. Lee G, Kim BS. Biological reduction of graphene oxide using plant leaf extracts. Biotechnology progress. 2014;30:463-469.

40. Rao S, Upadhyay J, Polychronopoulou K, Umer R, Das R. Reduced Graphene Oxide: Effect of Reduction on Electrical Conductivity. Journal of Composites Science. 2018;2:25.

41. Kuila T, Bose S, Khanra P, Mishra AK, Kim NH, Lee JH. A green approach for the reduction of graphene oxide by wild carrot root. Carbon. 2012;50:914-921.

42. Sykam N, Madhavi V, Rao GM. Rapid and efficient green reduction of graphene oxide for outstanding supercapacitors and dye adsorption applications. Journal of Environmental Chemical Engineering. 2018;6:3223-3232.

43. Hayyan M, Abo-Hamad A, AlSaadi MA, Hashim MA. Functionalization of graphene using deep eutectic solvents. Nanoscale research letters. 2015;10:324.

44. Wang J, Salihi EC, Šiller L. Green reduction of graphene oxide using alanine. Materials Science and Engineering: C. 2017;72:1-6.

45. Bagri A, Mattevi C, Acik M, Chabal YJ, Chhowalla… M. Structural evolution during the reduction of chemically derived graphene oxide. Nature …. 2010

46. Kaniyoor A, Ramaprabhu S. A Raman spectroscopic investigation of graphite oxide derived graphene. Aip Advances. 2012;2:032183.

47. Ruidíaz-Martínez M, Álvarez MA, López-Ramón MV, Cruz-Quesada G, Rivera-Utrilla J, Sánchez-Polo M. Hydrothermal Synthesis of rGO-TiO2 Composites as High-Performance UV Photocatalysts for Ethylparaben Degradation. Catalysts. 2020;10:520.

48. Hidayah NMS, Liu WW, Lai… CW. Comparison on graphite, graphene oxide and reduced graphene oxide: Synthesis and characterization. AIP Conference …. 2017

49. Thakur AK, Singh SP, Kleinberg MN, Gupta A, Arnusch CJ. Laser-Induced Graphene–PVA Composites as Robust Electrically Conductive Water Treatment Membranes. ACS applied materials & interfaces. 2019;11:10914-10921.

50. Radoń A, Włodarczyk P, Łukowiec D. Structure, temperature and frequency dependent electrical conductivity of oxidized and reduced electrochemically exfoliated graphite. Physica E: Low-dimensional Systems and Nanostructures. 2018;99:82-90.

51. Cai W, Piner RD, Stadermann FJ, Park… S. Synthesis and solid-state NMR structural characterization of 13C-labeled graphite oxide. …. 2008

52. Nakajima T, Matsuo Y. Formation process and structure of graphite oxide. Carbon. 1994

53. Meyer JC, Geim AK, Katsnelson MI, Novoselov KS, Booth TJ, Roth S. The structure of suspended graphene sheets. Nature. 2007;446:60.

54. Han Y, Gao C. Ultrathin Graphene Nanofiltration Membrane for Water Purification. Advanced Materials. 2013;23:3693-3700.

55. Nair RR, Wu HA, Jayaram PN, Grigorieva… IV. Unimpeded permeation of water through helium-leak–tight graphene-based membranes. Science. 2012

56. Boutilier MSH, Lee J, Chambers V, Venkatesh V, Karnik R. Water filtration using plant xylem. PloS one. 2014;9:e89934.